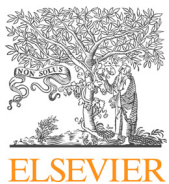
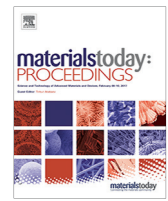

# A brief discussion on the tensile creep deformation behaviour of wrought single-phase γ-TiAl

Mainak Saha

*Department of Metallurgical and Materials Engineering, National Institute of Technology (NIT), Durgapur, India*

ABSTRACT

Creep deformation behaviour in single phase γ-TiAl alloy has been an extensively studied topic since the late 1970 s. A lot of literatures have reported creep behaviour of γ-TiAl alloys, manufactured using different processing techniques [1–7]. The present discussion revisits the original work on understanding the tensile creep deformation behaviour of wrought single-phase γ-TiAl alloy by Hayes et al. [8] and is aimed to develop an understanding of steady state creep, through strain vs strain rate and strain vs ln(strain rate) plots. Besides, it also attempts to study the variation of stress exponent with temperature between 760 and 900$^{o}$C and also, to determine activation energies using the two most common approaches, namely: Zener-Hollomon (Z-H) [9] and Sherby-Dorn (S-D, temperature compensated time approach) [10] for stress levels of 69.4 and 103.4 MPa between 760 and 900$^{o}$C.

## 1. Introduction

A series of reports on room temperature tensile deformation behaviour of γ- TiAl alloys show that the near- γ, two phase compositions having Al contents around 48 at.% possess the highest strengths and ductilities [1–3,11–18]. Extensive creep deformation studies have been carried out on a number of two phase near- γ TiAl alloys produced by various processing routes [8,19–29]. In addition, compression creep studies have been carried out on a number of binary single phase γ-TiAl alloys and a number of literatures have reported that minimum strain rates during creep testing at different regimes of temperature and stress, is hugely dependent on the grain size of materials [1,8,19,30,31]. Minimum strain rate of such intermetallic alloys may be defined in terms of Mukherjee-Bird-Dorn equation [32–34]. From a fundamental point of view, a study of the creep deformation behaviour of single phase TiAl is highly essential to provide useful insight into the deformation behaviour of the two phase TiAl alloys.

### 1.1. Motivation

The work of Hayes et al. [8] presents and discusses an analysis of the minimum strain rate deformation as well as an analysis of the tertiary creep behaviour of a wrought single phase γ-TiAl alloy within temperature range of 760-1000$^{o}$C and stress range of 32–345 MPa. The analysis by Hayes et al. [8] was carried out through determination of apparent creep activation energies, at different temperature regimes, for various stresses, followed by theoretical calculation of stress exponents at different stress regimes, for different temperatures. This was followed by the prediction of the main mechanism for creep rupture at different temperatures, in a given stress range, using Monkmann-Grant (M−G) plots [9] and extensive microstructural characterisation using Optical microscope and the Transmission Electron Microscope (TEM) [8].

The present discussion revisits the original work by Hayes et al [8] and is aimed to develop an understanding of steady state creep, through strain rate vs strain and ln(strain rate) vs strain plots. Besides, it also attempts to (i) determine variation in stress exponent with temperature between 760 and 900$^{o}$C and (ii) to determine activation energies through Zener-Hollomon (Z-H) and Sherby-Dorn (S-D, temperature compensated time approach) for stress levels of 69.4 and 103.4 MPa between 760 and 900$^{o}$C. Some of the major claims of the work by Hayes et al. [8] are:

*E-mail address:* mainaksaha1995@gmail.com

- Wrought single phase γ-TiAl alloy does not exhibit tertiary creep between 760 and 900 °C, at stress levels of 69.4 MPa and 103.4 MPa.
- There is no steady state creep observed at 832 °C, at stress levels of 69.4 MPa, for both interrupted and uninterrupted tests.
- γ-TiAl alloys do not exhibit dislocation creep at 760°C, 832°C and 900°C between stress level: 32–345 MPa.

## 2. Discussion

Table 1 shows the composition of single phase γ-TiAl alloy, forged, heat treated and finally subjected to tensile creep testing, as discussed in the work of Hayes et al [8]. As discussed earlier, that only constructing strain vs time plot to determine "min. strain rate" and designating the same to be the "steady state strain rate", may be highly misleading for a number of materials, which do not exhibit steady-state creep [18]. Ti-53Nb-1Al, for instance, whether tested to rupture or terminated at mid-strains of 0.18% and 0.5% at 832°C, exhibited a very less amount of steady-state creep. In the case of creep test till rupture, significant primary creep is exhibited whereas, in the case of creep test terminated at 0.18% strain, there is no primary creep regime and for the creep test till 0.5% strain, the primary creep regime is in between the two aforementioned extremes. In all cases, there is a very early onset of tertiary creep, However, the rate of tertiary creep in all the three samples is significantly different, as indicated by slopes of plots in Fig. 3 (a) and (b). Besides, from Fig. 1(a) and (b), the strain rate as well as strain for initiation of tertiary creep in three samples, also vary, to a small extent, with the sample having terminating strain of 0.5% showing tertiary creep initiated at the lowest strain rate and the highest strain.

**Table 1**
Composition of the alloy (used by Hayes et al. [8]).

| El.   | Ti   | Al   | Nb   | Fe    | H      | O      | N      | C      |
|-------|------|------|------|-------|--------|--------|--------|--------|
| at. % | 45.9 | 52.9 | 0.91 | 0.048 | 0.0219 | 0.1523 | 0.0026 | 0.0555 |

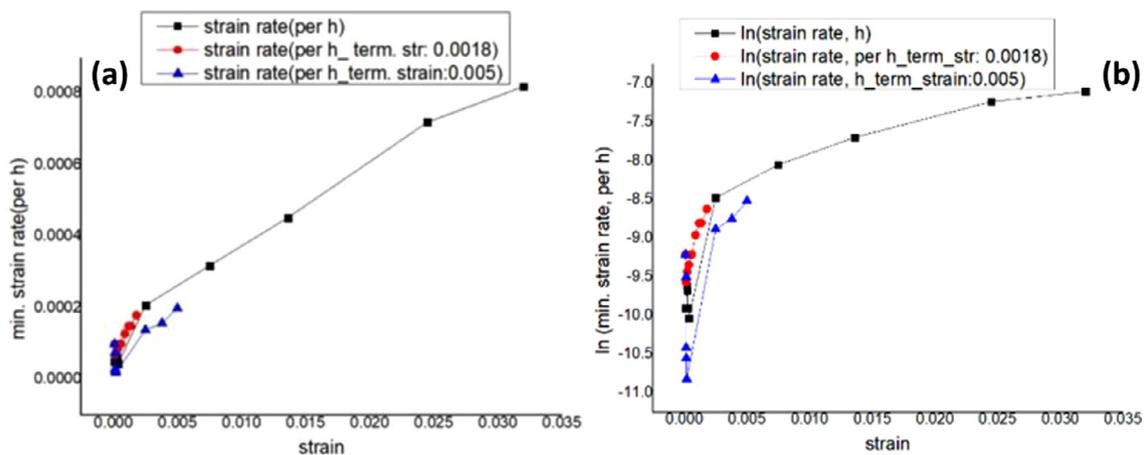

**Fig. 1.** (a) Strain rate vs Strain and (b) ln(strain rate) vs strain plots at 832 °C for sample subjected to interrupted tensile creep testing. 2 interruption (terminating) stresses were used by Hayes et al. [8] viz. 0.18% and 0.5% strain to fracture.

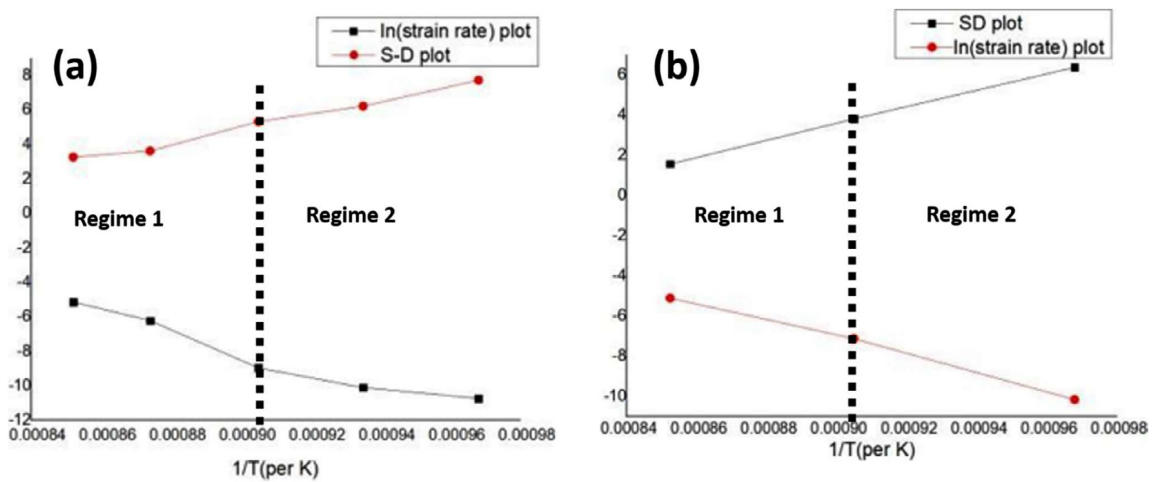

**Fig. 2.** ln (strain rate, in h) vs 1/T plot (marked as black) and Sherby -Dorn plot (marked as red) to determine creep activation energies at different temperature regimes at (a) 69.4 MPa and (b) 103.4 MPa. (For interpretation of the references to colour in this figure legend, the reader is referred to the web version of this article.)

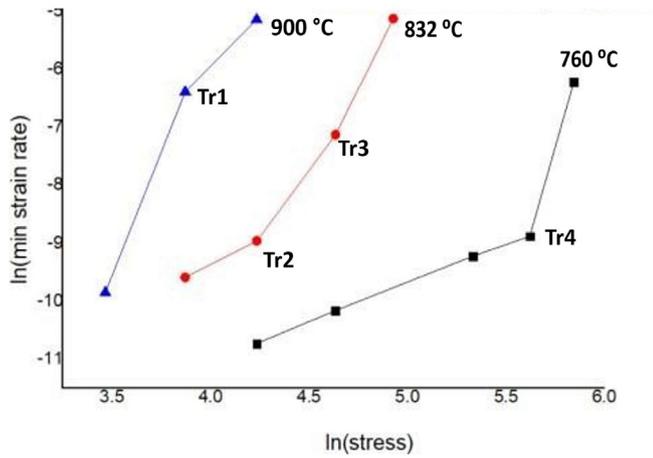

**Fig. 3.** ln(min strain rate) vs strain plots at 760, 832 and 900°C.

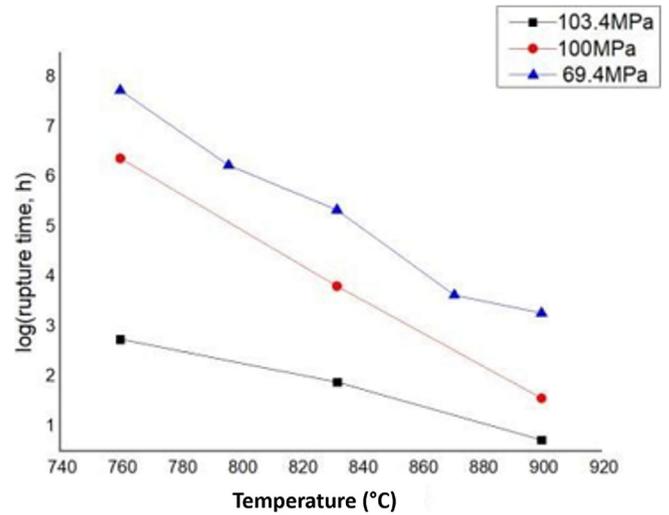

**Fig. 5.** Variation of log(rupture time, h) with temperature (°C) between 760 and 900°C, at applied creep stress levels of 69.4 (blue), 100 (red) and 103.4 MPa (black). (For interpretation of the references to colour in this figure legend, the reader is referred to the web version of this article.)

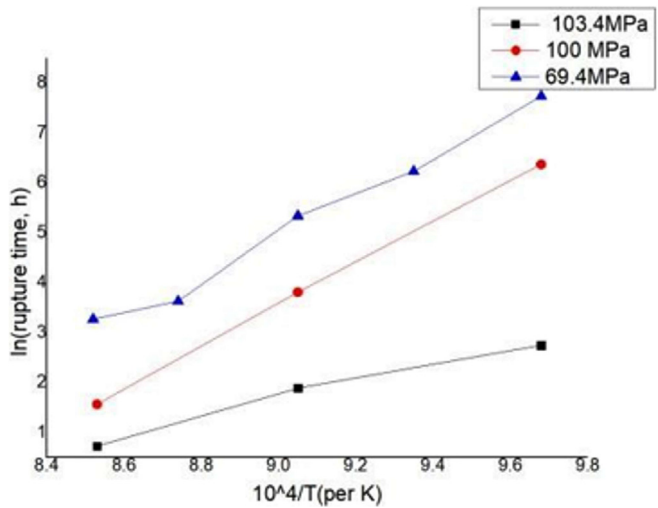

**Fig. 4.** Larson-Miller (L-M) plots at applied stresses of 69.4, 100 and 103.4 MPa.

### 2.1. Determination of activation energies between 760 and 1000°C at 69.4 and 103.4 MPa

#### 2.1.1. At 69.4 MPa

From Fig. 2(a), the apparent creep activation energies (determined from slope of black curve representing ln(strain rate, in h) vs 1/T at 69.4 MPa) vary from 423.92 kJ/mol in regime 1 to 409.82 kJ/mol in regime 2. Regimes 1 and 2 although seem to be independent [35], but the values of activation energies in two regimes, are found to be within experimental error. Moreover, the ZH (Zener Hollomon) parameter is obtained (from y-intercept of Fig. 2(a)) as 1.23e(14). Besides, Fig. 2(a) also shows rupture time vs 1/T plot (marked red) where SD parameter (determined from the y-intercept) is found to be 4.93e-14 and apparent creep activation energy is determined (from slope of Fig. 2) as 329.03 kJ/mol. The transition temperature from Regime 1 to 2 is ∼ 838.11 °C.

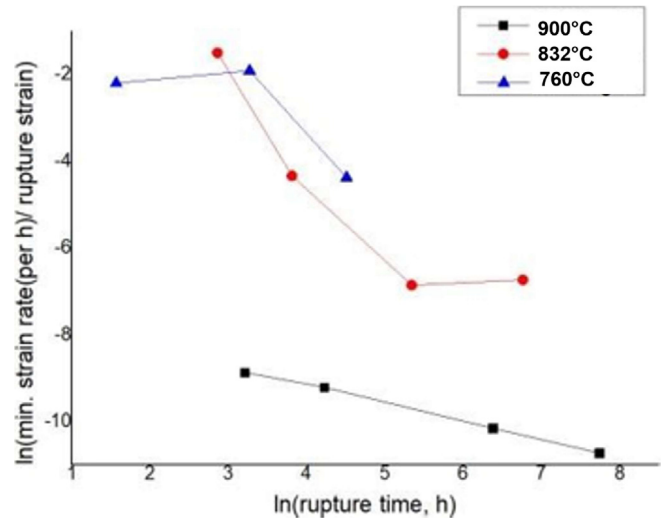

**Fig. 6.** Modified Monkman-Grant (M–G) parameter at 760, 832 and 900°C.

**Table 3**
Larson-Miller constants and parameters at 69.4, 100 and 103.4 MPa between 760 and 900°C.

| Stress (MPa) | L-M constant (K) | L-M Parameter |
|---|---|---|
| 69.4 | 1.74 | 14.05 |
| 100 | 3.96 | 30.64 |
| 103.4 | 4.15 | 34.02 |

**Table 2**
Grain boundary and lattice diffusion activation energies ($Q_{gb}$ and $Q_L$ (in kJ/mol), respectively) at different test temperatures between 760 and 900°C and creep stresses of 69.4 and 103.4 MPa following the work of Hayes et al [8]. $Q_{gb}$ and $Q_L$ have been calculated using Ashby's approach [9,37].

| Temperature (°C) | Stress (MPa) | Q (kJ/mol) | Qgb (kJ/mol) | QL(kJ/mol) |
|---|---|---|---|---|
| 760 | 69.4 | 192 | 72 | 120 |
| 760 | 103.4 | 304 | 114 | 190 |
| 832 | 69.4 | 560 | 210 | 350 |
| 832 | 103.4 | 405 | 151.88 | 253.13 |
| 900 | 69.4 | 624 | 234 | 390 |
| 900 | 103.4 | 519 | 194.63 | 324.38 |

**Table 4**
M–G plot vs modified M–G plot (considering linear fit throughout entire creep life) at 760, 832 and 900°C. where p, C and R² abbreviate for M–G exponent; M–G intercept and Goodness of fit, respectively.

| T (°C) | P (M–G) | C (M–G) | R² (M–G) | P (modified M–G) | C (modified M–G) | R² (modified M–G) |
|---|---|---|---|---|---|---|
| 760 | −1.4 | 1.43 | 0.96 | −1.56 | 1.61 | 0.97 |
| 832 | −1.3 | 1.31 | 0.93 | −1.33 | 1.40 | 0.98 |
| 900 | −0.4 | 0.52 | 0.99 | −0.68 | 0.69 | 0.99 |

*2.1.2. At 103.4 MPa*

From the ln(strain rate, in h) vs 1/T (per K) plot (marked as red in Fig. 2(b)), ZH parameter (determined from y-intercept) is 1.11e (14) and apparent creep activation energies (determined from slope of ln(strain rate, in h) vs 1/T plot) vary from 364.77 kJ/mol in regime 1 to 377.68 kJ/mol in regime 2. Regimes 1 and 2 are sequential [36].

Besides, Fig. 2(b) also shows rupture time vs 1/T plot (marked as black) where S-D parameter (Sherby- Dorn parameter, determined from y-intercept) is found to be 1.99e(−15) and Apparent Creep activation Energy (determined from slope) is found to be 345.65 kJ/mol. The transition temperature is decreased to ∼ 825.90°C about 14°C less than that at 69.4 MPa.

It is observed that at 832 and 900°C, with increase in stress level from 69.4 to 103.4 MPa, there is a decrease in apparent creep activation energy, but the reverse trend is observed at 760°C. This is subject to further investigations using microstructural investigation at 760°C, but is beyond the scope of discussion of the present work.

*2.2. Plots for determining stress exponent at 760, 832 and 900°C*

The plot (Fig. 3) for 760°C (black curve) shows that processes in 2 regimes must be independent. At Tr1: 244.69 MPa, the plot suggests that there is transition from dislocation glide controlled creep to dislocation climb controlled creep [36,38–43], suggesting a minor microstructural change leading to a change in creep deformation behaviour.

Based on the red coloured plot at 832 °C (Fig. 3): (i) regime 1 (from beginning to point Tr2) Stress exponent (from slope): 3; (ii) regime 2 (from points Tr2 to Tr3), Stress exponent: 4; (iii) regime 3 (from point Tr3 to end), Stress exponent: 4. Moreover, the red plot (in Fig. 3) also suggests that the cree mechanisms operating at three aforementioned regimes are independent. At Tr 2: 66.68 MPa there is transition from dislocation glide controlled creep to dislocation climb controlled which again suggests that there is change in mechanism in Dislocation creep due to minor microstructural changes.

Based on the blue coloured plot at 900 °C (Fig. 3): (i) regime 1 (from beginning to Tr4): stress exponent: 5 and (ii) regime 2 (from Tr4 to end): stress exponent: 4. This suggests that power law creep [44] is continued but with a different slope (stress exponent) which again suggests transition from dislocation climb to dislocation glide controlled creep and thus, no major microstructural change.

*2.3. Prediction of creep life using Larson-Miller (L-M) [45] and Manson-Haferd (M-H) [46] parameters*

From Fig. 4, it may be inferred that creep life of the alloy decreases, due to onset of creep rupture [45–47], with increase in stress between 69.4 and 103.4 MPa and temperatures between 760 and 900°C. Table. 3 shows Larson-Miller (L-M) constants and parameters at 69.4, 100 and 103.4 MPa between 760 and 900°C. It may be observed from Table 2. that with increase in stress levels from 69.4 to 103.4 MPa, both L-M constants and parameters increase between 760 and 900°C.

Based on linear interpolation from Fig. 5, the M–H parameter at Stresses: 69.4 MPa, 100 MPa and 103.4 MPa may be calculated to be equal to −0.033, −0.031 and −0.014, respectively.

*2.4. Prediction of creep life using modified Monkman-Grant (M–G) parameter [9,48–52]*

From Fig. 6 and Table. 4, it may be observed that predominant creep rupture mechanism is power law breakdown [18] at 760 and 832°C, due to p being greater than −1.4 but at 900°C, owing to p equal to −0.4, the predominant creep rupture mechanism requires further microstructural investigation and detailed analysis which is beyond the scope of the present study [53–55]. Besides, it may also be observed that using modified M–G plots, the goodness of fit is much improved than using the original M–G plots, implying that use of modified M–G plots improves data reliability [56].

**3. Conclusions**

Contradicting the claims of Hayes et al. [8], the various plots (Figs. 1-6), in the present discussion suggest that:

- Creep life between 760 and 900 °C, 69.4 MPa and 103.4 MPa is dominated hugely by tertiary creep.
- Hardly, any Steady state is observed at 832 °C, 69.4 MPa, for interrupted tensile creep tests.
- Dislocation creep tends to be the main deformation mechanism for these alloys at 760°C, 832°C and 900°C between 32 and 345 MPa.
- Creep life of the alloy decreases, due to onset of creep rupture, with increase in stress between 69.4 and 103.4 MPa and temperatures between 760 and 900°C.

**4. Future scope of the work**

A small attempt has been made to revisit the understanding in terms of creep behaviour in Ti-53Nb-1Al through strain rate vs strain plots, followed by determination of activation energies and stress exponents for different temperature and stress regimes, based on the work already performed by Hayes et al. [8]. However, extensive microstructural characterisation is required in order to study changes in microstructure, associated with variation in apparent creep activation energies in different temperature regimes and variation of stress exponent in different stress regimes. Besides, the effect of grain size on creep behavior of the alloy also needs to be studied in great details.

**CRediT authorship contribution statement**

**Mainak Saha**: Conceptualization, Data curation, Formal analysis, Funding acquisition, Investigation, Methodology, Project administration, Resources, Software, Supervision, Validation, Visualization, Writing - original draft, Writing - review & editing.

**Declaration of Competing Interest**

The authors declare that they have no known competing financial interests or personal relationships that could have appeared to influence the work reported in this paper.

**Acknowledgement**

MS would like to thank the Department of Metallurgical and Materials Engineering, NIT Durgapur, for detailed discussions during the scripting of the brief discussion.